\documentclass[hidelinks, 10pt, conference, letterpaper]{IEEEtran}

\usepackage{epsfig}
\usepackage{amssymb}
\usepackage{amsfonts}
\usepackage{verbatim}
\usepackage{graphicx}
\usepackage{url}
\usepackage{amsmath}
\usepackage{xcolor}
\usepackage[justification=centering]{caption}
\usepackage{subcaption}

\begin{document}
%

\title{Hydra: Leveraging Functional Slicing for Efficient Distributed SDN Controllers}

\author{\IEEEauthorblockN{Yiyang Chang\IEEEauthorrefmark{1},
Ashkan Rezaei\IEEEauthorrefmark{2},
Balajee Vamanan\IEEEauthorrefmark{2},
Jahangir Hasan\IEEEauthorrefmark{3},
Sanjay Rao\IEEEauthorrefmark{1} and
T. N. Vijaykumar\IEEEauthorrefmark{1}}
\IEEEauthorblockA{\IEEEauthorrefmark{1}School of Electrical and Computer Engineering, Purdue University \\ Email: \{chang256, sanjay, vijay\}@purdue.edu }
\IEEEauthorblockA{\IEEEauthorrefmark{2}Department of Computer Science, University of Illinois at Chicago \\ Email: \{arezae4, bvamanan\}@uic.edu}
\IEEEauthorblockA{\IEEEauthorrefmark{3}Google Inc. \\ Email: jahangir@google.com}
}


\maketitle

\begin{abstract}
The conventional approach to scaling Software-Defined Networking (SDN) 
controllers today is to partition switches based on network topology, with each 
partition being controlled by a single physical controller, running all SDN applications.
However, topological partitioning is limited by the fact that
(i) performance of latency-sensitive (e.g., monitoring) SDN applications 
associated with a given partition may be impacted by co-located 
compute-intensive (e.g., route computation) applications; (ii) simultaneously
achieving low convergence time and response times might be challenging;
and (iii) communication between instances of an application across partitions
may increase latencies. To tackle these issues, in this paper, 
we explore  {\em functional slicing}, a complementary approach to scaling,
where multiple SDN applications belonging to the same topological partition 
may be placed in physically distinct servers. We present \textit{Hydra}, a framework 
for distributed SDN controllers based on functional slicing. Hydra chooses
partitions based on convergence time as the primary metric, but places application
instances across partitions in a manner that keeps response times low while
considering communication between applications of a partition, and instances of
an application across partitions. Evaluations using the Floodlight controller
show the importance and effectiveness of Hydra in simultaneously keeping
convergence times on failures small, while sustaining higher throughput per
partition and ensuring responsiveness to latency sensitive applications.
\end{abstract}

\section{Introduction}
\label{sec:intro}

Software-Defined Networking (SDN) is becoming prevalent in datacenter and enterprise 
networks~\cite{b4,swan}. The central idea behind SDN is to consolidate 
control plane functionality (e.g., routing, access control) at a \textit{logically} 
centralized controller which monitors and manipulates network state~\cite{4d, openflow}. 
An SDN controller for a small network with hundreds of switches could
be hosted on a single physical server. However, as networks  grow in
size and functionality, the controller's compute and memory requirements
exceed one single server's capacity. 
Therefore, large datacenter and
enterprise networks distribute the controller functionality 
over multiple servers or VMs~\cite{elasticon,kandoo,hyperflow,onix}.

\begin{figure}
\centering
    \includegraphics[scale=0.25, trim=0cm 0cm 0cm 0cm, clip=true]{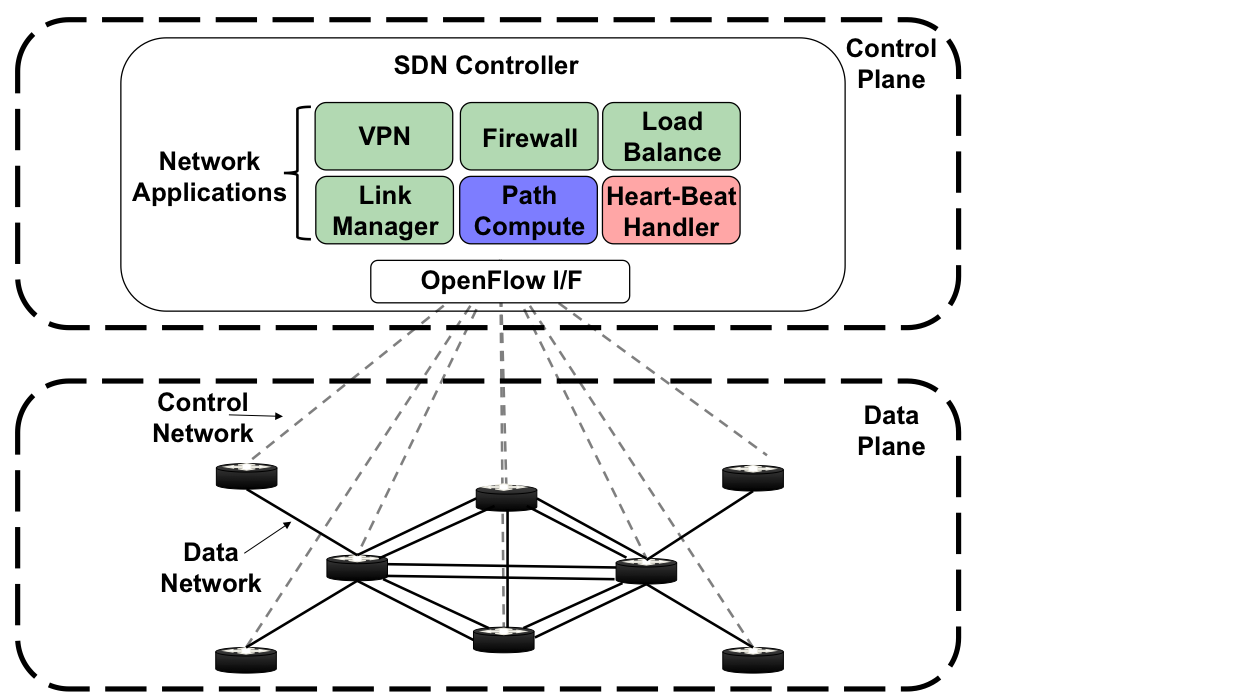}
\caption{\small{An example SDN network}}
\vspace{-0.3in}
\label{fig:SDN}
\end{figure}

Real SDN deployments typically consist of several tens of SDN applications 
for diverse network tasks such as routing, load-balancing, security, and 
Quality of Service (see Figure~\ref{fig:SDN}). Because these applications handle 
different events (e.g., link failure vs. path lookup) and perform diverse functions, 
they impose varying demands on the underlying machine resources. 
We broadly classify them into three groups:\\
(1) \textit{Real-time} applications that periodically refresh network state (e.g., 
link manager, heart-beat handler) expect a response within a timeout interval; 
failing to respond before deadline would trigger expensive false alarms (e.g., a spurious link failure).\\
(2) \textit{Latency-sensitive} applications that are invoked during flow setup (e.g., path lookup,
bandwidth reservation or QoS) are in the critical path of applications and directly impact 
flow completion times. Therefore, it is crucial to reduce their latency. 
However, they don't have a hard deadline constraint. \\
(3) \textit{Computationally intensive} applications such as shortest-path computation 
are triggered less often due to infrequent events such as link failures. 
But when triggered, these applications exert substantial pressure (load spikes) on
compute and memory. Convergence time, which is the time required for \textit{global}
state convergence (e.g., time required for find alternate paths in all 
partitions after a link failure in one partition), is an important metric for these 
applications. 

Designing a distributed control plane that scales well with network size and 
application heterogeneity is an important problem. 
The conventional approach to scaling SDN deployments~\cite{elasticon,kandoo,hyperflow,onix} 
is \textit{topological slicing} where the network topology is partitioned 
across multiple controller instances. Each controller instance, which runs 
on a single server machine, co-locates all applications and handles all events 
from a network partition containing a subset of switches. 

Topological slicing suffers from a few shortcomings:\\
(1) Because topological slicing co-locates all applications, 
finding the best partition size that satisfies all
applications (i.e., missed deadlines for real-time applications, 
latency for latency-sensitive applications, 
and convergence time for computationally intensive applications) 
is hard.
For instance, co-locating computationally intensive applications with other applications
may require smaller partition sizes (i.e., higher number of partitions) in order to satisfy 
resource constraints on the server machine. 
However, increasing the number of partitions would likely worsen convergence time for route 
recomputation on failures. 
Also, latency-sensitive applications such as bandwidth reservation and QoS 
may require communication across multiple
instances of the application running across partitions at flow setup time, potentially
leading to an increase in packet-in response times as the number of partitions increase.
Finally, there could be other administrative constraints on partition sizing
(e.g., a unit within an organization may want to have a separate controller instance).
In summary, while there is substantial diversity among applications, 
topological slicing is agnostic of the different applications' 
requirements, and, therefore, does not scale well. 
(2) Topological sizing hurts real-time and latency-sensitive applications.
Because computationally intensive applications are susceptible to load spikes, 
co-locating computationally intensive applications with real-time and latency-sensitive
applications adversely affects their latencies (real-time applications are most affected 
by co-location) as we show in figure \ref{fig:hb}. 

We propose \textit{functional slicing}, an approach that complements topological 
slicing by splitting different control-plane functions across multiple servers.
Functional slicing adds a new dimension to the partitioning problem and provides more 
freedom for placement of applications on different servers. 
With functional slicing, a switch may forward different events 
to different controllers (e.g., one could install a forwarding rule at the switch for each event
or have the original server forward the events to other servers).
With functional slicing, we can optimize the number of partitions to minimize
\textit{only} convergence time, without violating administrative constraints  
and without affecting real-time or latency-sensitive applications.

While functional slicing offers \textit{one more degree of freedom} for partitioning the control plane, 
it complicates placement. For instance, placing control-plane functions that are in the critical path 
in different machines would lead to longer flow completion times (i.e., the overhead of crossing 
machine boundaries would increase response times for packet-in messages during path setup). 
Therefore, our placement algorithm must be aware of the dependencies between the 
different control-plane functions. 

We present \textit{Hydra}, a framework for partitioning and placement 
of SDN control-plane functions in different servers. \textit{Hydra} leverages \textit{functional slicing} 
to increase flexibility in partitioning and placement. Moreover, Hydra's 
partitioning is \textit{communication-aware} -- Hydra considers the communication 
graph to avoid placing control-plane functions that are in the critical path 
in different machines. 
We first formulate
the placement of application instances across partitions as an optimization problem
with the objective of minimizing the latency of latency-sensitive applications that
are in the critical path, subject to resource constraints 
(i.e., number of servers, CPU and memory per server). 
We then reduce our formulation to a multi-constraint graph partitioning problem 
and solve it using well-known heuristics~\cite{karypis:mconstraint}.
To shield real-time applications from other applications, 
Hydra uses thread prioritization. 
Hydra assigns the highest priority to threads of real-time applications 
and next highest priority to latency-sensitive applications, 
while separating computationally intensive applications 
from the other two categories.  

\textit{Hydra} is relevant for both reactive controllers (where rules are installed after
examining the first packet of each flow), and proactive controllers (where rules are 
pre-installed in switches)~\cite{devoflow}.
Our optimization formulation is agnostic to the choice of the model. Our formulation 
considers the packet-in rates, which may be high for reactive  SDNs and low for pro-active SDNs,
and the rates, among other factors,  influence the best  partition chosen by \textit{Hydra}.
Our evaluation shows a range of packet-in rates to capture a continuum of this choice.

In summary, we make the following contributions:\\
$\bullet$ We propose \textit{functional slicing}, which adds a new dimension to 
partitioning and provides more flexibility. \\
$\bullet$ We introduce a \textit{communication-aware} placement algorithm that leverages 
functional slicing and avoids its potential shortcomings. \\
$\bullet$ We evaluate \textit{Hydra} using Floodlight~\cite{floodlight} controller 
and show the effectiveness of \textit{Hydra's} key techniques -- functional slicing, 
communication-aware placement, and prioritization.

The rest of the paper is organized as follows. Section \ref{sec:background}
presents an overview of Hydra's approach, and Section~\ref{sec:hydra} delves
into the details of Hydra. Section~\ref{sec:methodology} describes our
experimental methodology and Section~\ref{sec:results} presents our results.
Section~\ref{sec:relatedwork} discusses related work.  Finally, 
Section~\ref{sec:conclusion} concludes the paper.

\begin{figure*}[htp]
\hfill
\begin{subfigure}[t]{0.32\textwidth}
\centering
\includegraphics[scale=0.15, trim=0cm 0cm 0cm 0cm, clip=true]{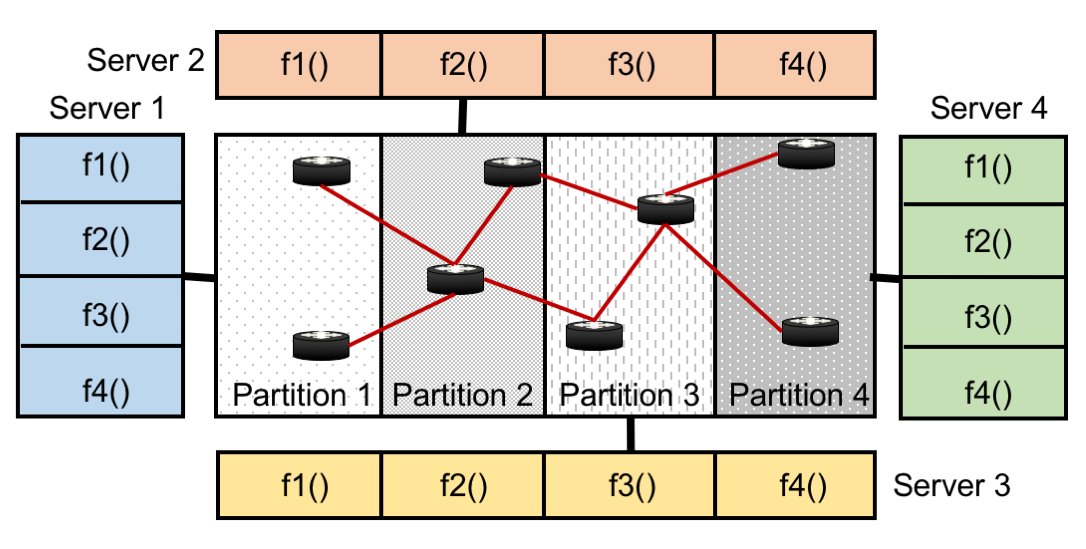}
\caption{Topological slicing}
\label{fig:topo-slice}
\end{subfigure}
\hfill
\begin{subfigure}{0.32\textwidth}
\centering
\vspace{-65pt}
\includegraphics[scale=0.15, trim=0cm 0cm 0cm 0cm, clip=true]{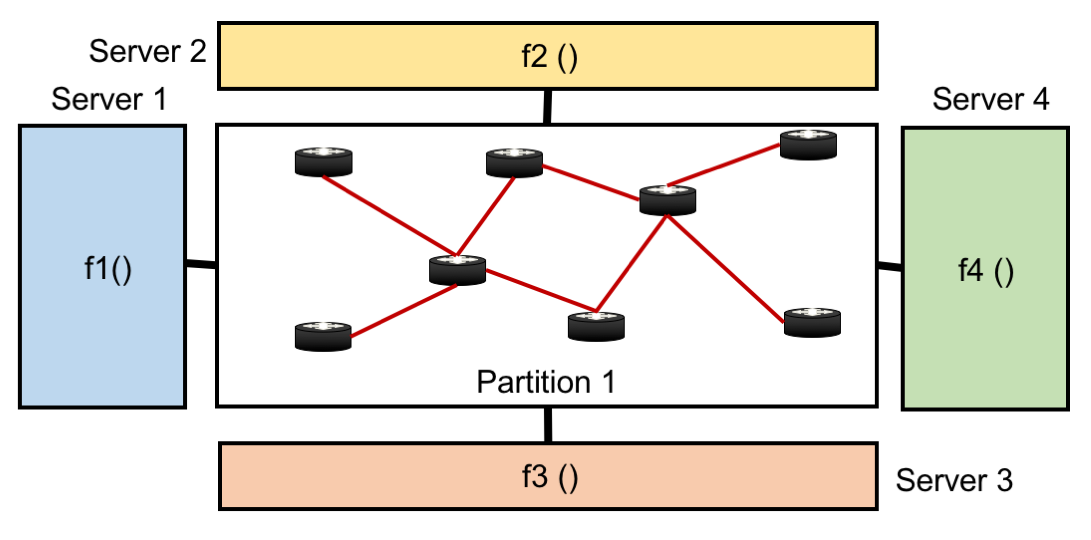}
\caption{Pure functional slicing}
\label{fig:func-slice}
\end{subfigure}
\hfill
\begin{subfigure}{0.32\textwidth}
\centering
\vspace{-65pt}
\includegraphics[scale=0.15, trim=0cm 0cm 0cm 0cm, clip=true]{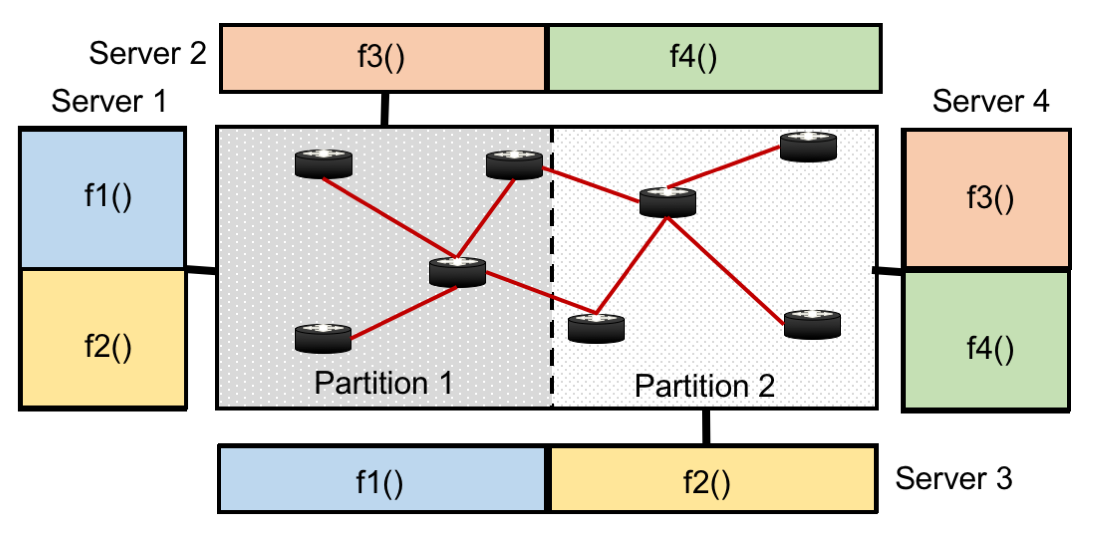}
\caption{Hybrid slicing with Hydra}
\label{fig:hybrid-slice}
\end{subfigure}
\caption{Approaches to partitioning controller functionality.}
\end{figure*}

\section{Hydra Rationale}
\label{sec:background}

We begin by discussing alternative ways to scale SDN controllers, and
present Hydra's approach and rationale:

\noindent
{\bf Topological partitioning:}
Current  distributed controllers \cite{elasticon,kandoo,hyperflow,onix},
de facto assume 
topological partitioning of the network into multiple controller domains, with one
controller instance per domain. Each controller instance runs all the
control-plane applications (e.g., topology modules, heart-beat handler that
monitors switch failures) but handles events only from the switches in its own partition.
Figure~\ref{fig:topo-slice} shows an example of topological partitioning
where each partition contains two switches and the four applications 
({\em f1} through {\em f4}) run on each controller.
While topological partitioning helps with scaling, the 
sustainable throughput is still limited by the fact that the 
compute and memory capabilities must be sufficient to handle all applications 
in that partition. Increasing the number of partitions to reduce partition sizes 
may not be feasible due to network administrator constraints and
since this may potentially increase route convergence time when recomputing paths 
on a switch or link failure. 
Finally, state changes in any partition of an application 
may need to be propagated to other partitions in order to maintain consistency 
of the application's global network state, and flow set up (e.g., for a QoS application)
may involve communications across application instances located in different partitions.

\noindent
{\bf Pure functional slicing:}
\textit{Functional slicing} partitions the control-plane functions 
belonging to the same topological partition and places the functions in different servers. 
Figure~\ref{fig:func-slice} shows an example of functional slicing for the same 
network as in Figure~\ref{fig:topo-slice}. The example shows the 
four functions {\em f1()} through {\em f4()} split across four controllers
each of which covers the entire network (i.e., all the four topological
partitions in Figure~\ref{fig:topo-slice}).  
While this tackles some of the issues with topological partitioning, the
sustainable throughput may now be bottlenecked by the most demanding 
application. 
Further, \textit{pure} functional slicing may worsen the latency to handle critical
packet-in events because the control-plane functions needed to handle each
such event may be spread across multiple machines 
(i.e., kernel overheads and networks delays would lie in the critical
path of packet-in event-handlers). 

\noindent
{\bf Hydra's approach:}
With Hydra, we explore a hybrid scheme that employs a combination of topological and functional
slicing to reduce both convergence times and packet-in
processing latencies. Figure~\ref{fig:hybrid-slice} shows an example of our hybrid slicing for 
the same network as in Figure~\ref{fig:topo-slice}. 
The example shows two topological partitions. Each controller  and two functional partitions of each of the 
topological partitions, so that only two servers for each function have to converge as opposed to the four servers in topological partition in Figure~\ref{fig:topo-slice}. At the same time, an event involving all four functions needs communication only between two servers as opposed to four servers in functional slicing in Figure~\ref{fig:func-slice}.

While Hydra separates computationally-intensive applications (i.e., path re-computation) from the other two categories, Hydra shields real-time applications (e.g., heart-beats) from latency-sensitive applications (e.g., path lookup) using \textit{thread prioritization}. Hydra assigns the highest priority to real-time applications and second highest priority to latency-sensitive applications.

\section{Hydra}
\label{sec:hydra}
In this section, we discuss Hydra's \textit{communication-aware} placement algorithm. 
Recall that Hydra leverages \textit{functional slicing} to
calculate the number of partitions that minimizes convergence time,
without negatively impacting real-time and latency-sensitive applications or
violating administrative constraints. 

\subsection{Finding the right partition size}
In the first step, we compute the 
number of partitions by considering \textit{only} the most critical computationally intensive
application that directly impacts convergence time on failures.
Often, the topology (route computation) application is the most critical application. 
While the exact number of partitions that minimizes convergence time is 
implementation dependent, in general, as we increase the number of partitions (starting from 1), 
the convergence time would decrease as the computation gets parallelized across partitions.
But, after some point, the communication overheads between parallel computing 
instances would start to overwhelm the benefits from parallelization. 
Thus, it is reasonable to expect a U-curve with the best partition size somewhere in the middle. 
But, Hydra's placement algorithm does not depend on 
the relationship between convergence time and the number of partitions. 

\subsection{Communication-aware placement: formulation}

\textit{Hydra} takes as input the different (topological) partitions of applications and their
demands (CPU and memory), resource constraints (i.e., CPU, memory, and number of servers), and the 
communication graph to calculate the best placement 
of the applications' partitions that minimizes latency. We assume that 
computationally-intensive applications (e.g., path computation) are isolated by 
placing those applications in separate machines (or VMs); 
simple prioritization might be sufficient in some cases as well. 
We cast placement of the applications' partitions 
as an integer linear programming (ILP) optimization problem. 
Because our problem is NP hard, we identify a efficient heuristic   
that can solve it in reasonable time.

Let $P$ be the number of topological partitions, 
$N$ the number of SDN applications deployed in the network, 
and  $S$ the number of physical servers dedicated for the SDN control-plane.
We want to bin-pack $P \times N$ application slices 
within $S$ server machines such that the average packet-in processing 
latency is minimized.

We represent the communication between the different application slices using 
a \textit{communication graph} whose vertices are application slices. 
Thus, there are $P \times N$ vertices 
in this graph. The edges in the graph denote communication between slices.
Communication can occur between two different applications in the same partition
(e.g., packets permitted by a firewall module may then be forwarded to a load-balancer),
as well as between two slices of the same application in different partitions
(e.g., a bandwidth reservation application between a source and destination in two
different partitions will require communication between the application slices
in the two partitions).

Let ${d}_{ij}$ denote the communication cost between two slices. 
Because we are interested in latency, the communication cost denotes 
the additional latency overhead if the slices are placed in different machines.
Let $A_i$ denote a vertex in the communication graph where $i \in [1,P \times N]$.
Then, depending on placement, we have 
the vector $F[i]=k$ which denotes that application slice $A_i$ is placed in machine $k$.

\subsubsection*{Objective function}

Next, we model latency of latency-sensitive events. 
Because these events typically traverse multiple application slices, event-handling latency would
depend on the total communication cost across these applications slices (i.e., path delay). 
Let $E = \{e_1,e_2,...,e_r\}$ be the events of interest, with their
associated paths, $\{p_1,p_2,..,p_r\}$, in the communication graph. Naturally, 
each path is a sequence of edges in the graph. 

Then, the cost of an event is given by:

\begin{equation}\label{eq_pathlat}
\begin{split}
	\forall p_{m} \in P,~t_{lat}(p_{m}) &= \sum\limits_{<i,j> \in p_m, F[i] 
           \neq F[j]} d_{ij}
\end{split}
\end{equation}

In this formulation, two slices would incur latency overhead of $d_{ij}$ 
when placed in different servers but no overhead when co-located in the same physical 
machine. 

We can assign a weight (e.g., relative priority, probability) to each event and 
calculate the weighted latency as follows. 

\begin{equation}\label{eq:tlat}
  t_{lat} = \sum\limits_{p_m \in \{p_1,p_2,\dots p_r\}} \gamma(p_m)t_{lat}(p_m)
\end{equation}

The weights could be relative priorities of the events 
based on semantic knowledge or could just be event probabilities. 
Our objective is to minimize equation (\ref{eq:tlat}) subject to capacity (i.e., 
CPU and memory), latency, and correctness constraints. 

\subsubsection*{Capacity constraints}
Let the compute and memory capacity of each 
server be $R_{cpu}$ and $R_{mem}$, respectively. Let $A_{i}$'s compute and 
memory requirements be $C_{i}$ and $M_{i}$, respectively. Then, we have 
the following constraints based on CPU and memory capacities. 

\begin{equation}\label{eq:contraints}
\begin{split}
  \underset{k}{\operatorname{max}}\left(\sum\limits_{\forall i: F[i]=k}{C_{i}} \right)
    &\leq R_{cpu} \\
  \underset{k}{\operatorname{max}}\left(\sum\limits_{\forall i: F[i]=k}{M_{i}} \right)
    &\leq R_{mem} \\
\end{split}
\end{equation}

\subsubsection*{real-time constraints}
We can bound the latency for real-time applications using an additional constraint of the form:

\begin{equation}\label{eq:hb}
  ~t_{lat}(p_{m}) <= deadline_{m}
\end{equation}

where $p_{m}$ is a path of a real-time event $m$ in the graph.

\subsection{Communication-aware placement: simplification}

The final form of the objective function $t_{lat}$ is the linear combination 
$t_{lat} = \sum_{F[i] \neq F[j]}\alpha_{ij}d_{ij}$, 
for some coefficients $\alpha_{ij}$. 
If we ignore the constraints (i.e., equations (\ref{eq:hb}) and (\ref{eq:contraints}), 
we see that $t_{lat}$ only depends on the weight of the edge-cut between the
partitions and our aim is to find such a mapping $F$. 
If we ignore only equation (\ref{eq:hb}), the problem reduces to
the well-known \textit{multi-constraint graph partitioning} \cite{karypis:mconstraint} problem.  
If each vertex $A_{i}$ is assigned a vector of weights $\langle C_{i},M_{i} \rangle$ denoting the compute 
and memory requirement of each slice, then the problem is \textit{equivalent} to 
finding a \textit{S-way} partitioning such that the partitioning satisfies a constraint 
associated with each weight, while attempting to minimize the weight of edge-cut.
Because multi-constraint graph partitioning is a known NP-hard problem \cite{karypis:mconstraint}, 
we employ heuristic methods from ~\cite{metis} 
which deliver high quality results in reasonable time.
While our heuristic solution ignores equation (\ref{eq:hb}),  
we did not observe appreciable degradation in our experiments.

\subsection{Discussion} \label{sec:form-other}
We discuss dynamic load adaptation and fault tolerance. 
\subsubsection*{Load adaptation}
Some previous papers (\cite{elasticon,pratyaastha}) argue for the controller's partitioning 
and placement to change according to instantaneous load from switches (e.g., packet-in rate). 
However, such dynamic re-partitioning and placement requires applications to 
re-partition and migrate their state which drastically affects controller performance and offsets the cost advantage 
of dynamic re-partitioning. This cost of reorganizing state applies to controllers that 
store state locally as well as to those that use a distributed datastore.
While controllers that store state locally must 
aggregate/split/migrate their state whenever partitioning/placement changes \cite{pratyaastha}, 
controllers that use a distributed datastore must reshard their datastore whenever the 
partitioning changes \cite{elasticon}. 
Because the cost of provisioning for the peak load is a small fraction (e.g., dedicating 100 servers 
for a 100,000-server datacenter is only 0.1\%) of total cost of 
ownership (TCO) of large datacenters, 
we provision enough servers to accommodate the peak load and do not change our partitioning
based on packet-in rate (load). Nevertheless, if desired, \textit{Hydra's} placement algorithm 
is fast enough to respond to load variations. 

\subsubsection*{Fault tolerance}
For fault tolerance reasons, it may be desirable to replicate SDN controllers
in each partition, either using a simple master-slave design for each partition,
or a more strongly consistent approach based on the Paxos algorithm~\cite{Paxos}.
While fault tolerance mechanisms are orthogonal to our work, 
it is easy to generalize \textit{Hydra} to handle the placement of replicas.
Specifically, a simple approach is to replicate the configuration
produced in the previous section as many times as needed for adequate fault tolerance. 
If it is also desirable to consolidate the number of physical controller machines,
our model could be extended by including additional variables for each replica,
and using the same placement algorithms described in the previous section. To
ensure that replicas of a given application/partition slice are not 
placed on the same physical host, additional constraints may be added
to require replicas be placed in different hosts.
Finally, there might be additional requirements that
parts of the network supplied by different power sources  need controller isolation
for fault tolerance. This constraint can be added to our formulation by requiring
that applications corresponding to these partitions not be co-located with each
other.

\section{Experimental Methodology}
\label{sec:methodology}
In this section, we present the details of our implementation and our 
evaluation methodology.

\textbf{SDN Applications:}
We use the Floodlight SDN controller ~\cite{floodlight}, which is a
widely used OpenFlow controller. We evaluate four control-plane
functions: 
\begin{enumerate}
\item \textit{Shortest path computation} (DJ): Shortest path computation based
on Dijkstra's algorithm, which runs whenever a new link (switch) is discovered or an
existing link (switch) fails.  
\item \textit{Firewall} (FW): Filters packet-in 
  messages based on a set of rules.
\item \textit{Route Lookup} (RL): Returns the complete path based on
source/destination pair in a packet-in header.
\item \textit{Heart-beat handler} (HB): Generates and forwards heart-beat
messages between switches and controllers; 
\end{enumerate}

DJ is a \textit{computationally intensive} intensive application;  
FW and RL are \textit{latency-sensitive} applications and 
are invoked during path setup; 
HB is \textit{real-time} application -- if a heart-beat is not 
processed within a deadline (i.e., heart-beat interval), 
a spurious link/switch failure would result which 
would trigger DJ. While a production SDN deployment would include 
tens of applications, it is hard for researchers to 
study a large number of applications at production scales.

\textbf{Load Generation}: 
\textit{Hydra's} evaluation requires large topologies 
with a few thousand switches. 
Because network emulators such as \textit{Mininet} model 
both control and data plane, 
they do not scale beyond a few tens of switches \cite{elasticon}.
Therefore, we use \textit{CBench} \cite{cbench}.
CBench generates packet-in events that stress
the control-plane without modeling a full-fledged data-plane.
While the current implementation of \textit{CBench} generates random packet-in 
messages (to potentially non-existent destinations), 
we modified CBench to generate packets that are meaningful to our topology.
We use a reactive model of SDN in our experiments. However, our results 
are generalizable to both pro-active or reactive models. 

\textbf{Topology:} 
Datacenters typically employ hierarchical topologies which provide high bisection bandwidth 
and good fault tolerance~\cite{fattree-amin,d2tcp,flowbender}. 
Our datacenter topology is a fat-tree with $2560$ switches. The topology is organized into 512 core
switches, and 32 pods, with each pod containing 32 Top of Rack (ToR) switches.

\section{Results} 
\label{sec:results}

In this section, we compare \textit{Hydra} to \textit{Topological slicing}
for the three types of applications.
Recall that we care about different metrics depending on the application type -- 
lower missed heart-beats (deadlines) for real-time applications (HB), 
lower latency (higher throughput) for latency-sensitive applications (FW,RL),
and lower convergence time for computationally-intensive applications (DJ).

We begin by showing how convergence time varies with the number of partitions 
which enables us to choose the right partition size. Then we show how 
our \textit{communication-aware placement} co-locates different application slices. 
Because our placement depends on CPU and memory utilization, 
we show CPU and memory utilizations which are sensitive to a
variety of parameters such as packet-in rates, topology sizes,
and other parameters. After placement, we compare 
missed heart-beats for HB 
and throughput (at near-saturation high loads, throughput is a \textit{proxy} for 
latency as queuing becomes the dominant latency component) for FW and RL. 

\begin{figure}
\centering
\includegraphics[scale=0.30]{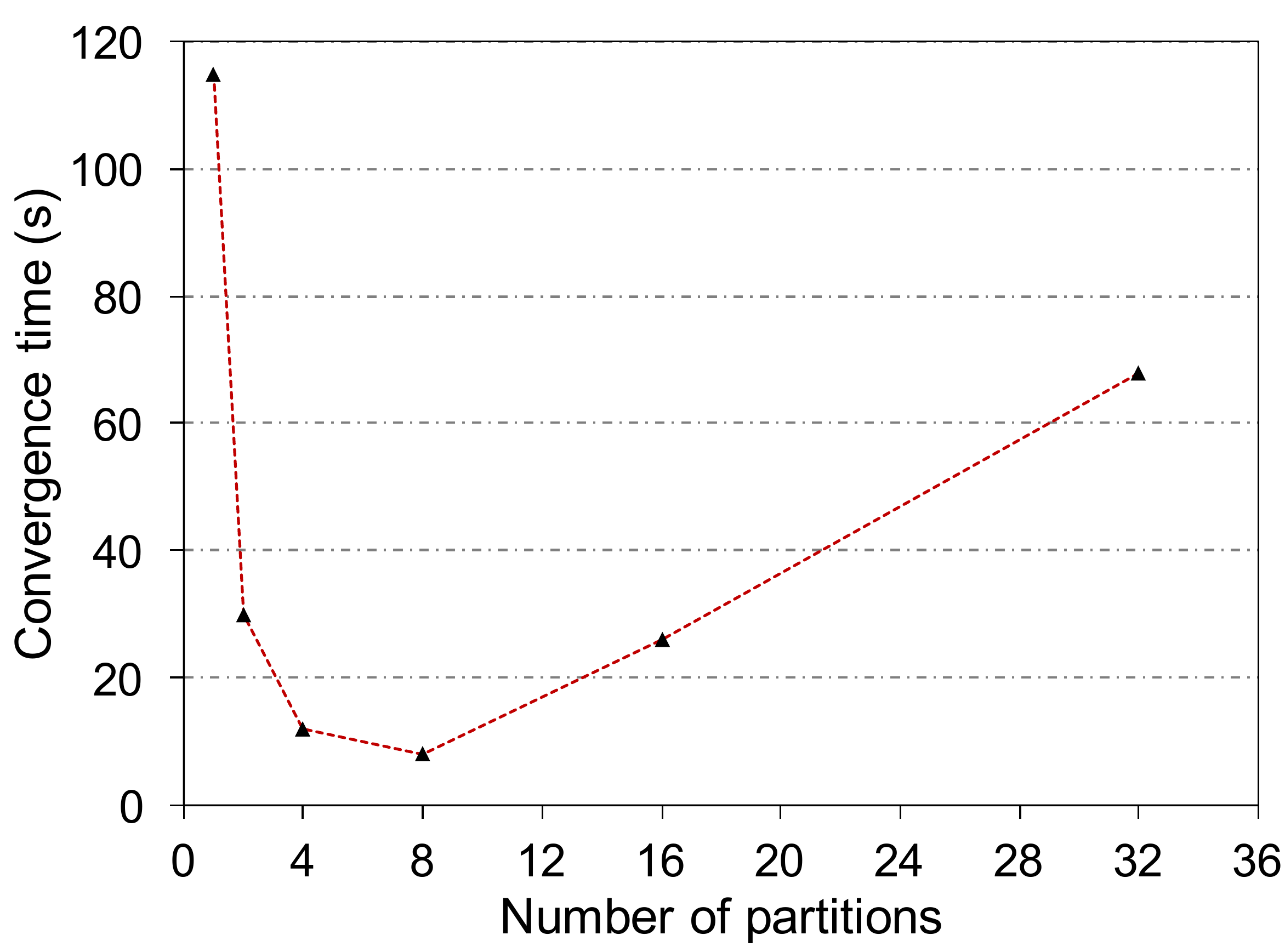}
\caption{Convergence time}\label{fig:curvefit}
\vspace{-0.2in}
\end{figure}

\subsection{Convergence Time}

We study convergence time for our fat-tree topology with 2560 switches. 
Because fat-tree is hierarchical, it is straight-forward to 
create partitions by grouping neighboring pods. 
For example, we can create two partitions by grouping 
16 pods in one partition and the other 16 in the other partition
(each pod contains 32 ToR switches). 
Recall that convergence time is the time to recalculate shortest paths 
after a link failure. So, to measure convergence time, we take down a \textit{random} link in 
our fat-tree which could be a border link (i.e., core link) 
or a partition-local link (i.e., ToR or aggregate links).  
We then measure the time required for \textit{all} partitions to 
recompute their paths which includes time for inter-partition communication.
While all neighboring partitions need to recompute on a border-link failure, 
a local link failure might also require partitions to advertise new costs 
to other partitions similar to BGP. 
For each partition size, we simulate 100 random link failures. 

We show the average convergence time for DJ vs. 
number of partitions (partition size) in figure \ref{fig:curvefit}.
We vary the number of partitions (ToR switches per partition) as 
1 (1024), 2(512), 4(256), 8(128), 16(64), and 32(32) along X-axis and 
show convergence time along Y-axis. We see that convergence time 
decreases \textit{rapidly} as we increase the number of partitions from 1 to 
8 due to amortization of compute from parallelization. 
However, after 8, convergence time starts to climb as 
communication overhead overwhelms gains from parallelization. 
Because topological slicing 
co-locates other applications with DJ, higher number of partitions are needed
to accommodate the aggregate CPU and memory requirements. In contrast, 
Hydra's \textit{functional slicing} enables us to choose the best 
partition size (e.g., 8 in this case), \textit{independent} of 
other applications.

\subsection{Communication-aware placement}

\begin{figure}
\centering
\includegraphics[scale=0.35]{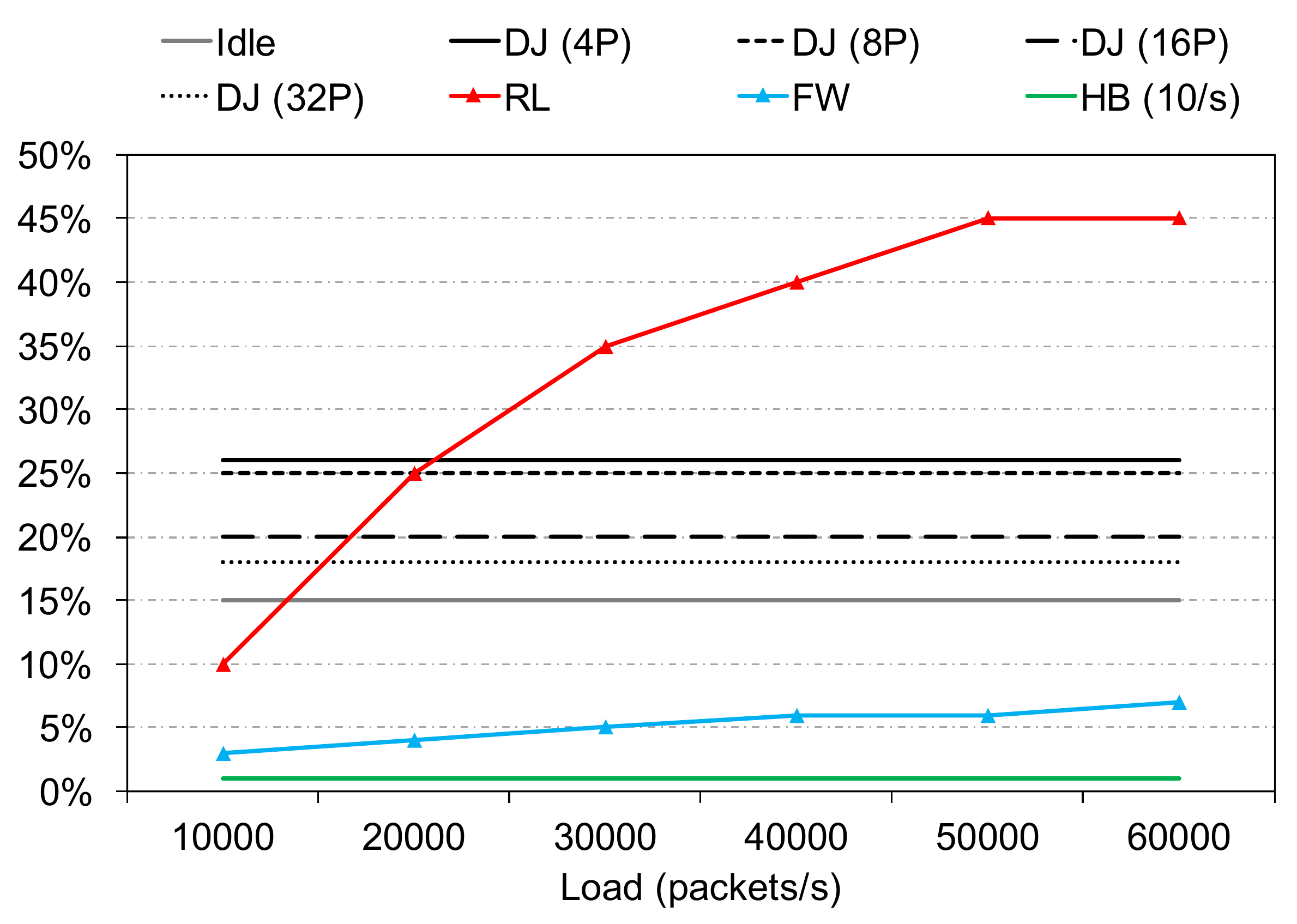}
\caption{Average CPU requirements}\label{fig:cpu}
\vspace{-0.2in}
\end{figure}

We start by showing the CPU and memory demands of applications. 
For these measurements, we ran Floodlight controller on our machine with  
$4$ cores of CPU and 64 GB of memory. 
The demand of each application depends on  
the amount of application state and controller's load.
Application state impacts both CPU and memory usage -- 
applications maintain state in memory and look up state 
for each packet-in message. 
RL must keep local topology information which depends on the partition size. 
The number of firewall rules impacts FW's state overhead.
In our experiments, we use $50,000$ firewall rules which is typical for large networks.
DJ maintains both local and global topology information. 
DJ's CPU usage depends on link failure rate and partition size. 
We simulate a random link failure every $10$ seconds which is reasonable for 
large networks.
From figure \ref{fig:curvefit}, we expect that 
DJ's CPU usage to be \textit{highly} sensitive to partition size. 
HB's CPU and memory usage are minimal -- its CPU usage slowly 
grows with heart-beat frequency but negligible overall.

The CPU demands of applications also depend on load (i.e., 
rate at which the controller receives packet-in messages from switches). 
We modified \textit{CBench} to precisely control packet rate. 
Our base controller saturates around 50,000 packets per second. 
Therefore, we make measurements from 10,000 to 50,000 packets per second. 
Even without any applications, SDN controllers 
run some common functions (e.g., south-bound \textit{OpenFlow} protocol handlers)
which cannot be turned off.
Therefore, we initially measure the idle CPU and memory usage without any applications (no 
incoming packets to the controller) which represents the overhead of 
starting a new controller instance. The overhead is about $15\%$ CPU usage and 
$512 MB$ of memory. 
We enable applications one-by-one and measure CPU and memory usage for each application 
(excluding idle overhead) at $100~ms$ intervals.
We discard initial and final samples to capture \textit{steady-state} usage.

Figure \ref{fig:cpu} shows the CPU requirements of different applications as 
as we vary the load. DJ and HB do not depend on load -- DJ's CPU usage depends on 
partition size and link failure rate (1 every $10$ seconds), and HB's usage depends on  
heart-beat frequency (we ran HB at 10/second and 100/second 
but they are both insignificant). We show DJ for varying partition sizes --
for example, DJ(4P) is for 4 partitions each with one fourth the number of switches as DJ(1P). 
We observe that DJ's CPU usage reduces with increasing number of partitions due to 
reduced number of switches. 
As discussed in the previous section, with topological slicing,
the state overheads of other applications (e.g., RL, FW) determine the partition size 
which negatively impacts convergence time. For instance, we can see that the 
combined CPU usage of \textit{Idle, RL, FW, HB, and DJ(4P)} is close to $100\%$ 
($15+45+7+3+25$) for higher loads. 
In fact, only when there are more than 8 partitions, the combined CPU usage
falls well below $100\%$ (servers usually operate at less than $90\%$ loads
to provide reasonable response times). 
Therefore, topological slicing is forced to choose a partition size of 16 or more 
which leads to high convergence times (see figure \ref{fig:curvefit}).
\textit{Hydra}, on the other hand, separates DJ from other applications, enabling
DJ to use the \textit{best} partition size. 

Memory usage is largely independent of load. 
Table \ref {table:mem} shows the average memory overheads of DJ, FW, and RL
for the one partition case containing all switches. 
From the table, it is clear that memory does not impact our
placement in our controller as all of applications comfortably
fit within our memory capacity.
However, we expect production controllers to have large state
overheads that will not fit within one server's memory.
We do not show  HB's memory overhead as it is negligible. 

\begin{table}[ht]
\caption{Memory requirements}
\centering
\begin{tabular}{| c | c | c |} 

\hline
\rule{0pt}{3ex} \textbf{DJ} & \textbf{RL} & \textbf{FW} \\ [0.5ex]
\hline

\rule{0pt}{3ex} 6.25 GB & 3.75 GB & 1.25 GB \\ [0.5ex]
\hline

\end{tabular}
\label{table:mem}
\end{table}

Recall from section \ref{sec:methodology} that our communication graph has only one edge 
between RL and FW, as RL and FW are the only applications that lie in the critical path 
of flow's path setup; DJ and HB do not have edges between them
or to either RL or FW. 
From figure \ref{fig:cpu} and table \ref{table:mem}, it is straight forward 
to see the difference between Topological slicing's and Hydra's placement decisions. 
Topological partitioning requires 16 controller instances (16 partitions) requiring 
16 cores. Each instance would host all the applications. 
In contrast, \textit{Hydra} creates 8 network partitions 
(\textit{minima} in figure \ref{fig:curvefit}). 
For each partition, it assigns two controller instances which run on separate CPU cores. 
While one controller instance hosts DJ for that partition, another instance hosts
all the \textit{other} applications -- RL, FW, and HB. 
While we could manually calculate optimal placements in this simple controller, 
deployment-scale controllers would likely consist of tens of applications with complex 
communication patterns, and, therefore, would require a rigorous approach such as Hydra.
Unfortunately, it is harder for researchers to experiment with production-scale controllers 
without access to production-scale networks and workloads. 

\subsection{Latency-sensitive applications}

\begin{figure}
\centering
\includegraphics[scale=0.35]{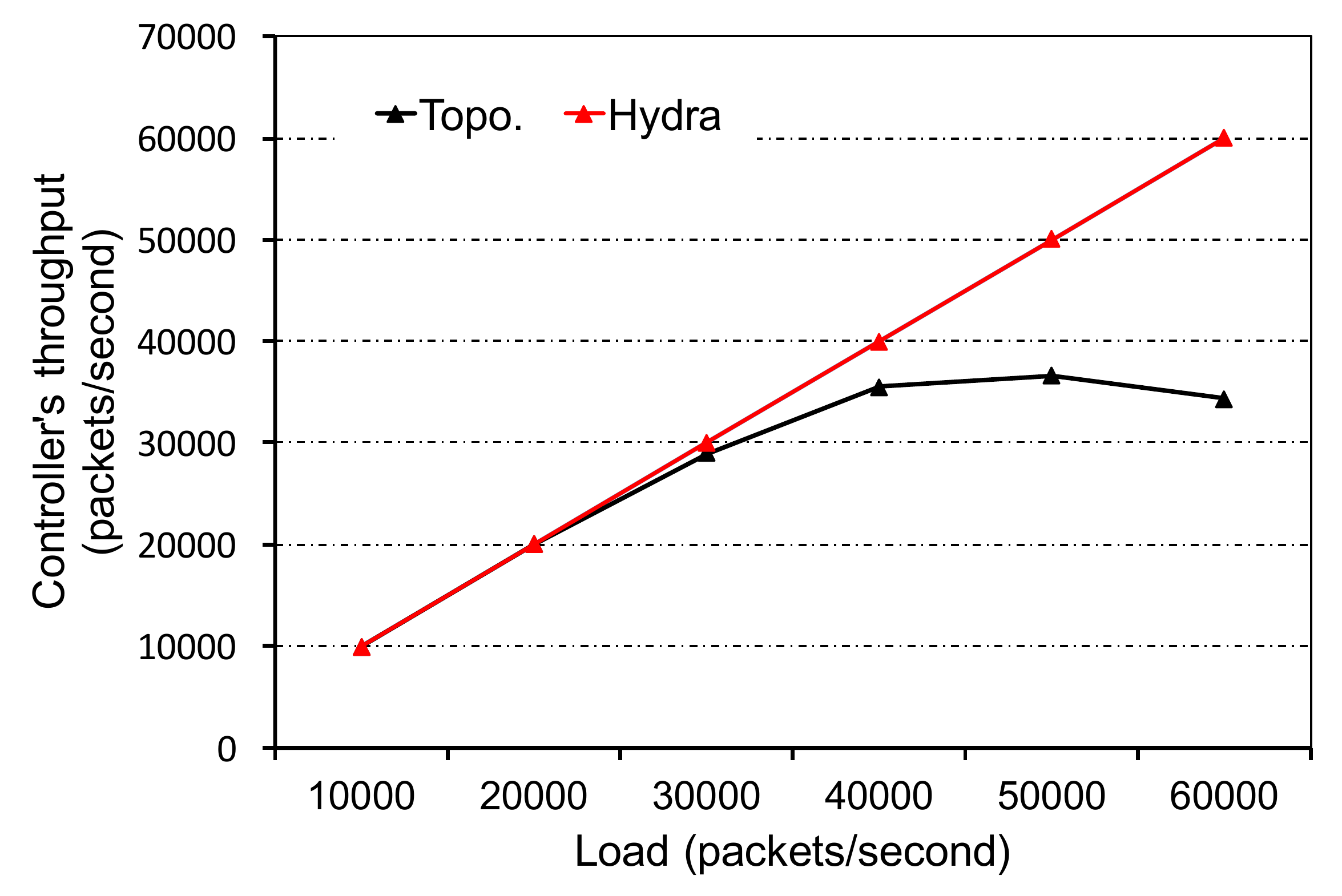}
\caption{Scalability of latency-sensitive applications in Hydra}\label{fig:scale}
\vspace{-0.2in}
\end{figure}

In this experiment, we compare the performance of latency-sensitive applications 
in \textit{one network partition}. Recall that Hydra creates 8 network partition
($1/8^{th}$ switches) as opposed to topological slicing which creates 16 partitions 
($1/16^{th}$ switches). In figure \ref{fig:scale}, we compare the 
scalability of latency-sensitive applications in Hydra vs. topological slicing. 
We show load (injected packets per second) along X-axis and 
the achieved throughput after route lookup (RL) and firewall processing (FW)
along Y-axis.
As we can see, Hydra scales well beyond $60,000$ packets per second 
whereas topological slicing saturates at about $40,000$. 
As a result, latency-sensitive events incur high queuing inside the controller 
in the case of topological slicing. 
It is also interesting to note that even though Hydra handles events from a larger number of switches,
the latency-sensitive applications (RL and FW) are isolated from the load spikes 
caused by computationally-intensive DJ application, thanks to \textit{functional slicing}.

\subsection{Real-time applications}

\begin{figure}
\centering
\includegraphics[scale=0.30, trim={0.15in 0.15in 0in 0.25in},clip]{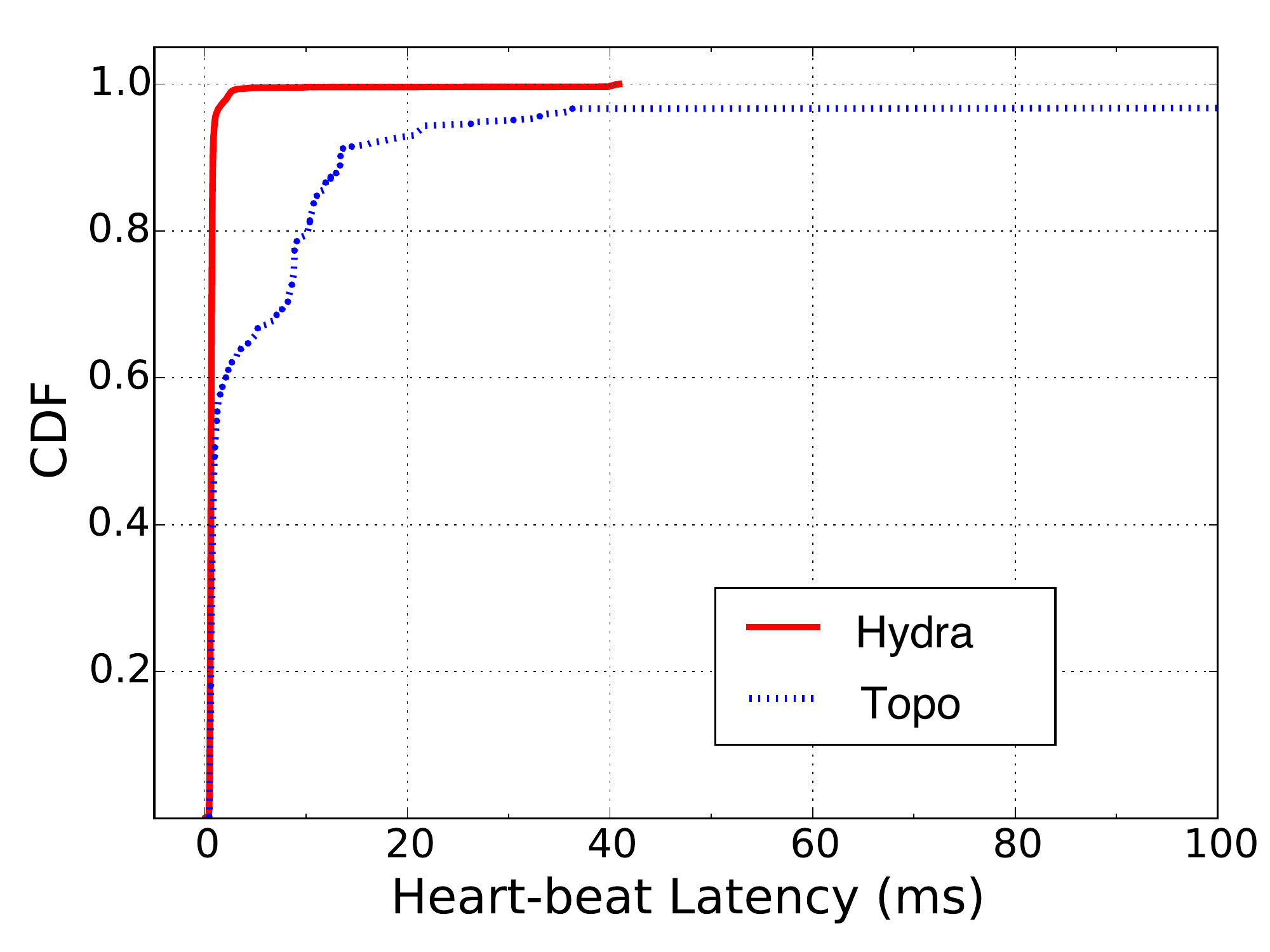}
\caption{Performance of real-time apps. in Hydra} \label{fig:hb}
\vspace{-0.2in}
\end{figure}

Separating computationally-intensive DJ application also helps our real-time heart-beats (HB) application. 
Figure \ref{fig:hb} shows the CDF of heart-beat latency between Hydra and topological slicing. 
Our default heart-beat frequency is 10 heart-beats per second. 
We see a marked difference between the two -- while \textit{Hydra's} 
$95^{th}$ and $99^{th}$ \%-iles are about $10$ ms,  
topological slicing's $95^{th}$ \%-ile is about $30$ ms. 
With a deadline of $100~ms$ (i.e., periodicity of heart-beats), 
topological slicing would suffer about $3\%$ missed deadlines, 
whereas Hydra would not miss \textit{any}.
While $3\%$ may look like a small number, but penalty for missed deadlines 
is very high (i.e., missed deadlines trigger expensive path recomputation 
which would further exacerbate the problem).

\subsection{Isolating the impact of prioritizing}

\begin{figure}
\centering
\includegraphics[scale=0.30, trim={0.15in 0.15in 0in 0.25in},clip]{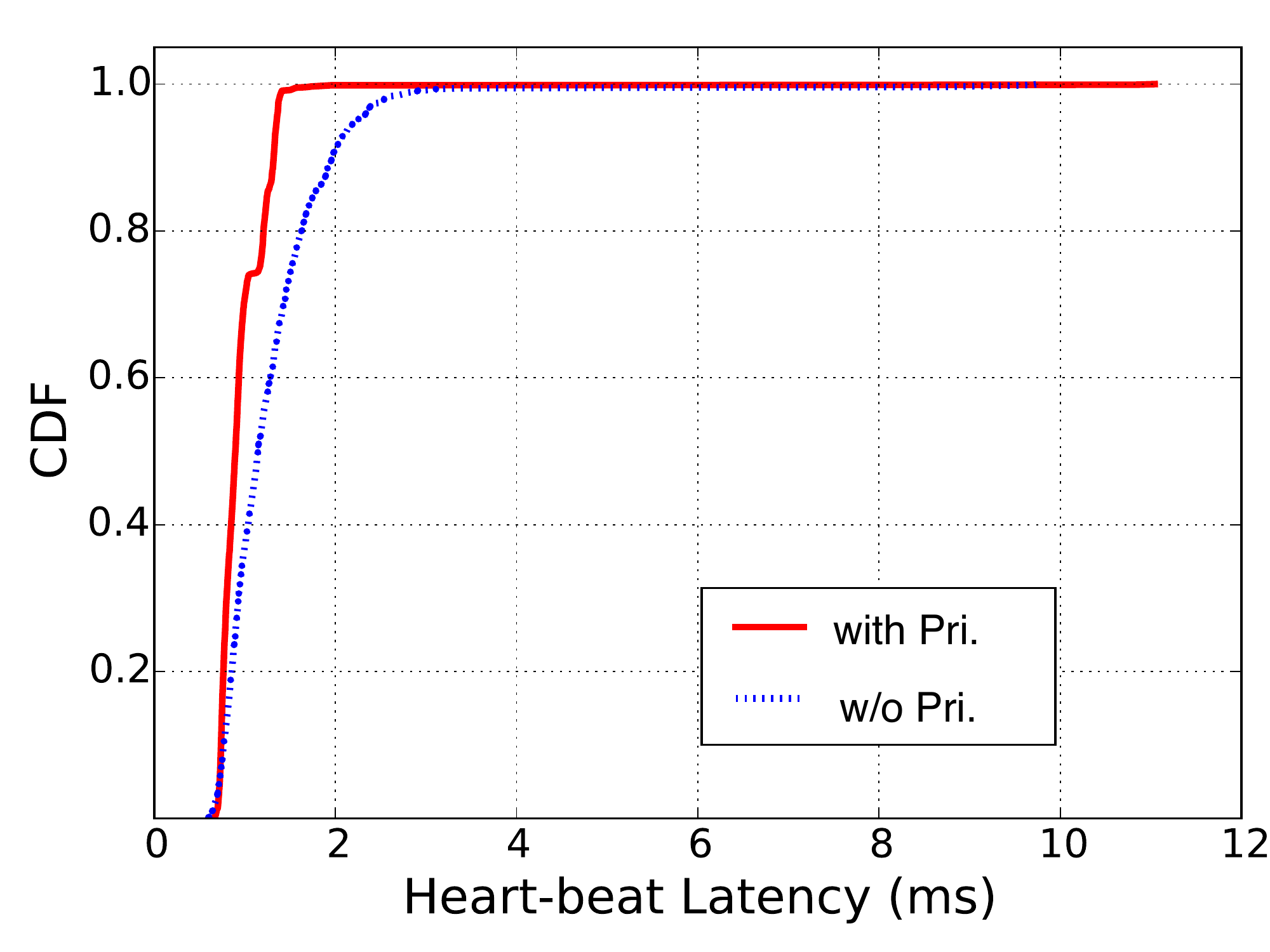}
\caption{Isolation of prioritization's gains} \label{fig:hb-pri}
\vspace{-0.2in}
\end{figure}

\begin{figure}
\centering
\includegraphics[scale=0.30, trim={0.15in 0.15in 0in 0.25in},clip]{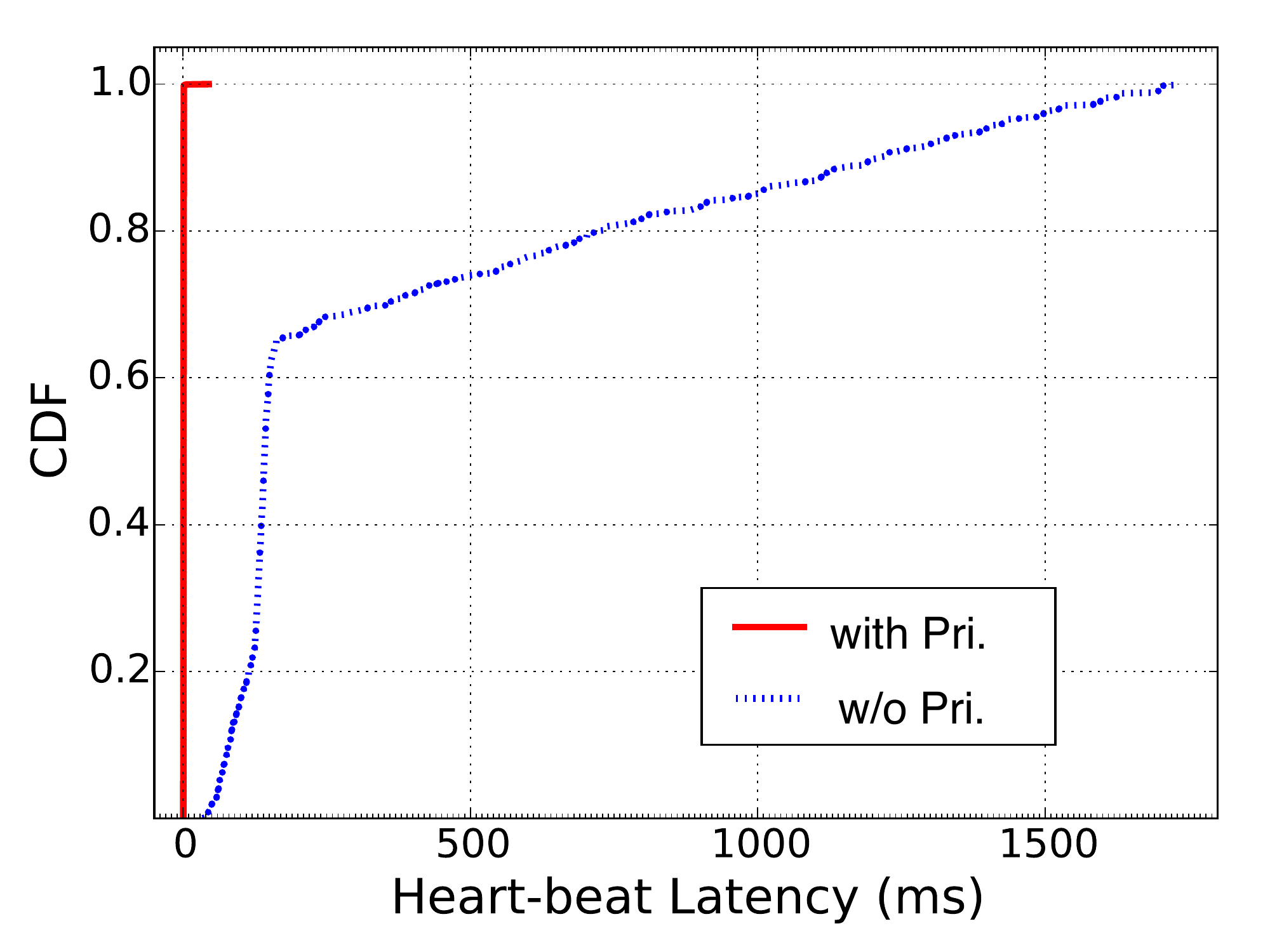}
\caption{Sensitivity to heart-beat rate} \label{fig:hb-pri-100}
\vspace{-0.2in}
\end{figure}

In this section, we isolate the gains from prioritizing real-time applications over 
latency-sensitive applications. In figure \ref{fig:hb-pri}, we compare the 
CDF of heart-beat latency between Hydra with and without prioritization.
The responses are received in a timely fashion when
HB is prioritized over RL, but modestly degrades when not prioritized. 
In figure \ref{fig:hb-pri-100}, we increase the heart-beat rate to 
$100$ per second to facilitate quicker failure detection. 
We see that \textit{almost} all HB messages meet the deadline when prioritized
but \textit{no} messages meet the deadline when not prioritized. 
In fact, some HB messages take as long as $1800~ms$ to get a response.
Thus, prioritization improves timeliness of real-time applications 
beyond functional slicing.

\section{Related Work}
\label{sec:relatedwork}
While there is a plethora of research on SDN, 
a systematic analysis of controller partitioning and placement is not well-studied.
Onix \cite{onix} focuses on providing APIs for control-plane and state distribution. 
Beehive \cite{beehive} enables applications to express their 
state-dependence and uses the inferred state-dependence to co-locate functions \textit{within} each application. 
In contrast, Hydra considers event-processing pipeline across applications and considers others constraints 
(e.g., CPU load, memory) to partition applications as well as the state (i.e., topology). 

Hyperflow \cite{hyperflow} improves controller performance 
by pro-actively synchronizing state but does not deal with partitioning.
Kandoo \cite{kandoo} offloads switch-local events to switches but does not address a large subset
of events that are not local to the switch.
ElastiCon \cite{elasticon} topologically partitions the controller based on CPU load. 
In contrast, Hydra employs a hybrid of topological and functional partitioning. 
A few other papers address the placement of the controller on the
network to reduce network delays and to topologically-slice the network for
better performance \cite{devolved,placement}. But none of them employ functional slicing and 
they do not target specific response times and convergence costs. 
While some papers \cite{onix,elasticon} argue for a logically-separate, globally-consistent, 
distributed datastore for storing state to ease communication among different 
controllers, others \cite{pratyaastha} prefer that the state be distributed among 
controller instances like many distributed or parallel applications today.
Nevertheless, our optimization formulation is agnostic to the choice of state management.  
In our evaluation,  we use Floodlight \cite{floodlight} which assumes the latter alternative
where there is no separate datastore but other communication costs (e.g., datastore) can be 
easily incorporated into our model.

\section{Conclusion}
\label{sec:conclusion}
In this paper, we have presented \textit{Hydra}, a 
framework for distributing SDN control functions 
across servers. \textit{Hydra} combines well-known topological slicing
with our novel \textit{functional slicing} and distributes applications based on 
their communication pattern. We have demonstrated
the importance of functional slicing and \textit{communication-aware} placement
in the scalability of SDN with extensive evaluations.

Our results, while promising, are only a start. First, while we evaluated
using applications that are available publicly controllers, 
we expect \textit{Hydra's} benefits to be even higher
with large-scale deployments. 
Getting access to production SDN deployments can enable larger-scale evaluations, which is an interesting 
direction for future work. Second, we are building a more comprehensive
system based on functional slicing, that can handle other issues such
as incrementally placing applications as loads drastically change and 
incorporating consistency guarantees into the model.

\bibliographystyle{IEEEtran}
\bibliography{IEEEabrv,sigproc}  

\begin{thebibliography}{10}
\providecommand{\url}[1]{#1}
\csname url@samestyle\endcsname
\providecommand{\newblock}{\relax}
\providecommand{\bibinfo}[2]{#2}
\providecommand{\BIBentrySTDinterwordspacing}{\spaceskip=0pt\relax}
\providecommand{\BIBentryALTinterwordstretchfactor}{4}
\providecommand{\BIBentryALTinterwordspacing}{\spaceskip=\fontdimen2\font plus
\BIBentryALTinterwordstretchfactor\fontdimen3\font minus
  \fontdimen4\font\relax}
\providecommand{\BIBforeignlanguage}[2]{{%
\expandafter\ifx\csname l@#1\endcsname\relax
\typeout{** WARNING: IEEEtran.bst: No hyphenation pattern has been}%
\typeout{** loaded for the language `#1'. Using the pattern for}%
\typeout{** the default language instead.}%
\else
\language=\csname l@#1\endcsname
\fi
#2}}
\providecommand{\BIBdecl}{\relax}
\BIBdecl

\bibitem{b4}
S.~Jain, A.~Kumar, S.~Mandal, J.~Ong, L.~Poutievski, A.~Singh, S.~Venkata,
  J.~Wanderer, J.~Zhou, M.~Zhu, J.~Zolla, U.~H\"{o}lzle, S.~Stuart, and
  A.~Vahdat, ``B4: Experience with a globally-deployed software defined wan,''
  in \emph{Proceedings of the ACM SIGCOMM}.\hskip 1em plus 0.5em minus
  0.4em\relax ACM, 2013, pp. 3--14.

\bibitem{swan}
C.-Y. Hong, S.~Kandula, R.~Mahajan, M.~Zhang, V.~Gill, M.~Nanduri, and
  R.~Wattenhofer, ``Achieving high utilization with software-driven wan,'' in
  \emph{Proceedings of the ACM SIGCOMM}, 2013, pp. 15--26.

\bibitem{4d}
A.~Greenberg, G.~Hjalmtysson, D.~A. Maltz, A.~Myers, J.~Rexford, G.~Xie,
  H.~Yan, J.~Zhan, and H.~Zhang, ``A clean slate 4d approach to network control
  and management,'' \emph{SIGCOMM Comput. Commun. Rev.}, vol.~35, no.~5, pp.
  41--54, 2005.

\bibitem{openflow}
N.~McKeown, T.~Anderson, H.~Balakrishnan, G.~Parulkar, L.~Peterson, J.~Rexford,
  S.~Shenker, and J.~Turner, ``Openflow: Enabling innovation in campus
  networks,'' \emph{SIGCOMM Comput. Commun. Rev.}, vol.~38, no.~2, pp. 69--74,
  2008.

\bibitem{elasticon}
A.~A. Dixit, F.~Hao, S.~Mukherjee, T.~Lakshman, and R.~Kompella, ``Elasticon:
  An elastic distributed sdn controller,'' in \emph{Proceedings of the ANCS},
  2014, pp. 17--28.

\bibitem{kandoo}
S.~Hassas~Yeganeh and Y.~Ganjali, ``Kandoo: A framework for efficient and
  scalable offloading of control applications,'' in \emph{Proceedings of the
  HotSDN}, 2012, pp. 19--24.

\bibitem{hyperflow}
A.~Tootoonchian and Y.~Ganjali, ``Hyperflow: A distributed control plane for
  openflow,'' in \emph{Proceedings of INM/WREN}, 2010, pp. 3--3.

\bibitem{onix}
T.~Koponen, M.~Casado, N.~Gude, J.~Stribling, L.~Poutievski, M.~Zhu,
  R.~Ramanathan, Y.~Iwata, H.~Inoue, T.~Hama, and S.~Shenker, ``Onix: A
  distributed control platform for large-scale production networks,'' in
  \emph{Proceedings of OSDI}, 2010, pp. 1--6.

\bibitem{karypis:mconstraint}
G.~Karypis and V.~Kumar, ``Multilevel algorithms for multi-constraint graph
  partitioning,'' in \emph{Proceedings of the {ACM/IEEE} Conference on
  Supercomputing, {SC} 1998, November 7-13, 1998, Orlando, FL, {USA}}, 1998,
  p.~28.

\bibitem{devoflow}
A.~R. Curtis, J.~C. Mogul, J.~Tourrilhes, P.~Yalagandula, P.~Sharma, and
  S.~Banerjee, ``Devoflow: Scaling flow management for high-performance
  networks,'' in \emph{Proceedings of the ACM SIGCOMM}, 2011, pp. 254--265.

\bibitem{floodlight}
``Floodlight,'' \url{http://www.projectfloodlight.org}.

\bibitem{metis}
G.~Karypis and V.~Kumar, ``A fast and high quality multilevel scheme for
  partitioning irregular graphs,'' \emph{SIAM J. Sci. Comput.}, vol.~20, no.~1,
  pp. 359--392, Dec. 1998.

\bibitem{pratyaastha}
A.~Krishnamurthy, S.~P. Chandrabose, and A.~Gember-Jacobson, ``Pratyaastha: An
  efficient elastic distributed sdn control plane,'' in \emph{Proceedings of
  the HotSDN}.\hskip 1em plus 0.5em minus 0.4em\relax New York, NY, USA: ACM,
  2014, pp. 133--138.

\bibitem{Paxos}
L.~Lamport, ``Paxos made simple,'' \emph{ACM Sigact News}, vol.~32, no.~4, pp.
  18--25, 2001.

\bibitem{cbench}
``Controller benchmark,''
  \url{http://www.openflowhub.org/display/floodlightcontroller/Cbench}.

\bibitem{fattree-amin}
M.~Al-Fares, A.~Loukissas, and A.~Vahdat, ``A scalable, commodity data center
  network architecture,'' in \emph{Proceedings of the ACM SIGCOMM 2008}, 2008,
  pp. 63--74.

\bibitem{d2tcp}
B.~Vamanan, J.~Hasan, and T.~Vijaykumar, ``Deadline-aware datacenter tcp
  (d2tcp),'' in \emph{Proceedings of the ACM SIGCOMM 2012}, 2012, pp. 115--126.

\bibitem{flowbender}
A.~Kabbani, B.~Vamanan, J.~Hasan, and F.~Duchene, ``Flowbender: Flow-level
  adaptive routing for improved latency and throughput in datacenter
  networks,'' in \emph{Proceedings of CoNEXT}, 2014, pp. 149--160.

\bibitem{beehive}
S.~H. Yeganeh and Y.~Ganjali, ``Beehive: Towards a simple abstraction for
  scalable software-defined networking,'' in \emph{Proceedings of
  HotNets-XIII}, 2014, pp. 13:1--13:7.

\bibitem{devolved}
A.-W. Tam, K.~Xi, and H.~Chao, ``Use of devolved controllers in data center
  networks,'' in \emph{INFOCOM WKSHPS}, April 2011, pp. 596--601.

\bibitem{placement}
B.~Heller, R.~Sherwood, and N.~McKeown, ``The controller placement problem,''
  in \emph{Proceedings of HotSDN}, 2012, pp. 7--12.

\end{thebibliography}

\end{document}